\documentclass[11pt]{article}
\input{epsf}
\usepackage{graphicx}
\usepackage{subcaption}
\usepackage{multirow}
\usepackage{comment}
\usepackage{indentfirst}







\begin{document}
\addtolength{\baselineskip}{0.01in}

\vspace*{0.5cm}

\begin{center}
{\large\bf
Optimization of Spatial Dose Distribution for Controlling Sidewall Shape in Electron-beam Lithography
}
\end{center}

\vspace{1mm}
\begin{center}
Pengcheng Li\\
Department of Electrical and Computer Engineering \\
Auburn University, Auburn, AL, USA \\
\end{center}

\bigskip

\begin{center}
{\bf Abstract}
\end{center}
Electron-beam (e-beam) lithography is widely employed in fabrication of 2-D patterns and 3-D structures.  A certain type or shape of the sidewall in the remaining resist profile may be desired in an application, e.g., an undercut for lift-off and a vertical sidewall for etching, or required for a device. Also, as the feature size is decreased well below a micron, a small variation of the sidewall slope can lead to a significant (relative) CD error in certain layers of resist.  Therefore, it is important to understand effects of spatial dose distribution on sidewall shape and be able to achieve the desired shape.  In this study, via simulation, the relationship among the total dose, spatial distribution of dose, developing time and sidewall shape, and performance of the method developed to optimize the dose distribution for a target sidewall shape have been analyzed.  The simulation results have been verified through experiments.

\vspace{6cm}

\setcounter{page}{1}
\clearpage
\addtolength{\baselineskip}{0.19in}

\section{Introduction}

Two-dimensional (2-D) patterns and three-dimensional (3-D) structures are transferred onto a resist layer via electron-beam (e-beam) lithographic process in various applications, e.g., discrete devices, photomasks,molds for imprint lithography, etc.  One of the critical factors which determine if the overall fabrication process is successful is the remaining resist profile after development.  A certain type of resist profile is desired depending on the subsequent process following resist development.  For example, an undercut profile is preferred for lift-off and a straight vertical sidewall for etching.  Also, as the feature size is reduced down to nanoscale, the aspect ratio of developed feature in the resist profile becomes larger even for a thin resist.  This makes a small variation in the sidewall slope cause a relatively large critical-dimension (CD) error.  Therefore, it is important to have a sufficient control over the sidewall shape in the resist profile.

The sidewall shape obtained through e-beam lithographic process depends on factors such as exposure (energy deposited in resist) distribution, developing time, etc.  Varying developing time is a passive approach in that the spatial exposure distribution is set, and therefore has a limited controllability.  Controlling the exposure distribution, more precisely the 3-D distribution of exposure, enables a more explicit method to achieve a target sidewall.  Nevertheless, in most of the previous work, only the dose level was varied with a uniform dose within a feature, to achieve different shapes of sidewall, and the 3-D exposure distribution was not considered \cite{Murali2006}-\cite{Klimpel2011}.  Changing the level of uniform dose only scales the exposure distribution without altering the spatial distribution and therefore does not fully utilize the available controllability of exposure distribution.  In a previous work \cite{Dai2011_1,Dai2011_2}, a 3-D exposure model was introduced to analyze effects of beam energy, resist thickness, feature size, and developing time on the spatial exposure distribution, i.e., depth-dependent proximity effect.  Subsequently, the issue of 3-D proximity effect correction was addressed, deriving the spatial dose distributions required for three different types of resist profiles using a simple search method, and only the simulation results were presented \cite{Dai2012,Zhao2014}.

In this study, a general-purpose optimization method, Simulated Annealing (SA), is adopted in determining the dose distribution required for target sidewall shapes and a number of experiments have been carried out to verify the simulation results obtained in the previous and this studies.  It is known that the SA is capable of finding the globally optimal solution (dose distribution) via a stochastic search process.  It also allows changing the doses of multiple regions in each iteration of optimization.

The paper is organized as follows. The exposure and developing rate models, and the development simulation are described in Section \ref{sec:3M}.  The proposed scheme of optimizing the dose distribution is described in detail in Section \ref{sec:SC}. Simulation and experimental results are provided along with discussion in Section \ref{sec:RD}, followed by a summary in Section \ref{sec:SM}.

\section{3-D Model}
\label{sec:3M}

\subsection{Exposure distribution}

\begin{figure}[!htb]
 \centering
 \epsfxsize=8cm
 \ \epsfbox{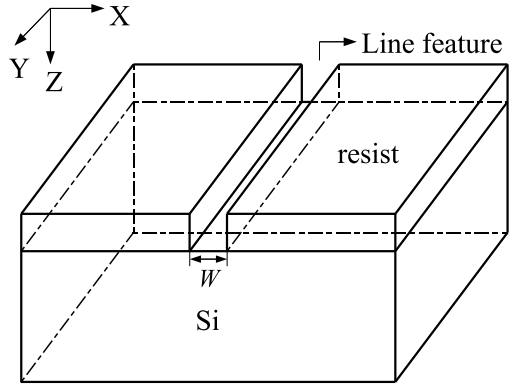} \\
 (a) \\
 \epsfxsize=8cm
 \ \epsfbox{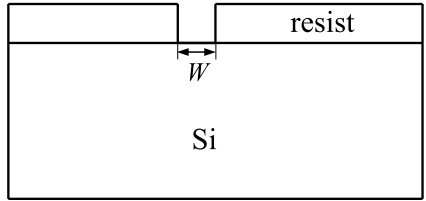} \\
 (b) \\
\vspace{0.1cm}
 \caption{(a) A line pattern transferred onto resist and (b) the cross-section of the remaining resist profile.}
\label{fig:cross}
\end{figure}

Consider a resist layer on a substrate where an X-Y plane corresponds to the top surface of the resist layer and the Z-axis is along the resist depth dimension, pointing down, as shown in Fig. \ref{fig:cross}(a).  Let $d(x,y,0)$ represent the e-beam dose at the point $(x,y,0)$ on the surface of the resist. The point spread function (PSF) \cite{Nilsson2011}-\cite{Lee2011}, which describes the distribution of deposited energy throughout the resist layer when a point on the top surface of resist is exposed, is denoted by $psf(x,y,z)$. Then the 3-D exposure distribution $e(x,y,z)$ in the resist layer can be expressed as follows:

\vspace{-0.5cm}
\begin{equation}
e(x,y,z) = \int\int d(x-x',y-y',0)psf(x',y',z)dx'dy'
\label{eqn:e}
\end{equation}

\vspace*{0cm}

In this study, the sidewall of resist profile for a line pattern is considered as illustrated in Fig. \ref{fig:cross}(a).  When the line pattern is sufficiently long (along the Y-dimension in Fig. \ref{fig:cross}(a)), the exposure can be assumed not to vary along the Y-dimension.  For such a line, it is sufficient to analyze only the cross-section of resist profile in the middle of line, perpendicular to the length dimension as shown in Fig. \ref{fig:cross}(b).  In the rest of this paper, only the cross-section plane (X-Z plane) is considered.  Then, the exposure distribution is denoted by $e(x,z)$.

\subsection{Developing Rate}

The relationship between the exposure and resist-developing rate is nonlinear.  Let the relationship be represented by a non-linear function $F[~]$ to be referred to as (exposure-to-developing-rate) {\em conversion formula}.  Then, $r(x,z)=F[e(x,z)]$ where $r(x,z)$ is the developing rate at point $(x,z)$.  At the center of a line (in the cross-section plane)where the exposure is highest when a uniform dose is given to the line, resist development progresses mainly downward (i.e., along the vertical dimension) such that the lateral development may be ignored.  Exploiting this property, the conversion formula is derived using a part of the 3rd-order polynomial curve.  It models the cross-section of resist layer by a 2-D array of blocks within each of which the exposure and therefore developing rate are assumed to be constant.  Using the remaining resist profile from experiments, the average developing rate can be computed for the vertical column of blocks at the center of line, and the initial conversion formula is derived from the average developing rate.  Then, the conversion formula is calibrated iteratively by remodeling the developing rate block by block.

 The following conversion formula was obtained:

\vspace{-1cm}
\begin{equation}
r = -2.485 \times 10^{-30}\times e^3+1.499 \times 10^{-19}\times e^2+2.201\times 10^{-8}\times e
\label{eqn:r}
\end{equation}

where $e$ is in eV/$\mu$m$^3$  and $r$ is in nm/min.

\subsection{Simulation of Resist Development}

In the earlier study, the resist development was simulated by the cell removal method \cite{Hirai1991}.  The computationally-intensive nature of the cell removal method makes the dose calculation procedure extremely time-consuming since the development simulation needs to be carried out many times through iterations.  Also, it often results in rough resist profiles.  A new simulation method was recently developed, which is orders of magnitude faster than the cell removal method and generates smooth profiles \cite{Dai2014}.  The overall shapes of profiles obtained by the new method are equivalent to the respective profiles by the cell removal method. It first considers only vertical development and subsequently all possible developing paths consisting of lateral development following vertical development.  The new simulation scheme was employed in this study though it will be presented in a future paper due to the page limit.

When a small region has a much higher exposure than its surrounding regions, its effective developing rate is significantly lower than the nominal rate (given by Eqn. \ref{eqn:r} due to the aspect-ratio-dependent development).  To reflect this effect in development simulation, the developing rate is adjusted according to the spatial distribution of exposure before the simulation.

\section{Sidewall Control}
\label{sec:SC}

\begin{figure}[!htb]
 \centering
 \epsfxsize=8cm
 \ \epsfbox{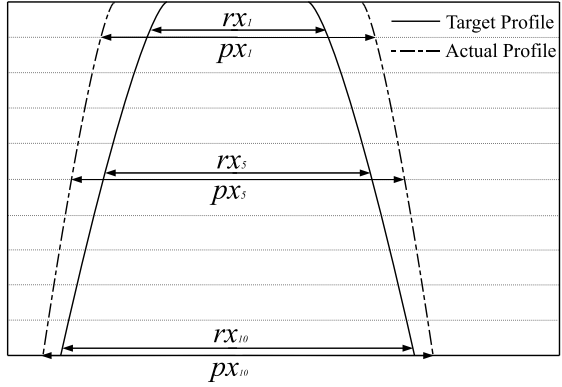} \\
 \caption{An illustration of sidewall-shape specification in the cross-section: $rx_i$ and $px_i$ are the target and actual widths of line in the $ith$ layer of resist, respectively. The cost function is defined as $C=max_i(|rx_i-px_i|)$.}
\label{fig:critialPoint}
\end{figure}

Given a developer and a developing time, the resist profile depends on the exposure distribution $e(x,z)$.  Therefore, one may attempt to control $e(x,z)$ in order to achieve a target resist profile.  When a substrate system is given, $e(x,z)$ is determined by the distribution of e-beam dose within the feature, i.e., a line.  The e-beam dose is varied (controlled) only along the width dimension, i.e., X-axis and therefore the dose distribution is denoted by $d(x)$.  The feature considered in this study is a long line and the cross-section of resist profile at the center of the line is characterized by the line widths in the top, middle and bottom layers as illustrated in Fig. \ref{fig:critialPoint}.  Let $rx_i$ and $px_i$ represent the target and actual widths in the $i$th layer, respectively.  Then, the optimization problem for sidewall control can be defined as finding $d(x)$ such that the cost function $max_i(|rx_i-px_i|)$ is minimized.

\vspace*{0cm}

\begin{figure}[!htb]
 \centering
 \epsfxsize=13cm
 \ \epsfbox{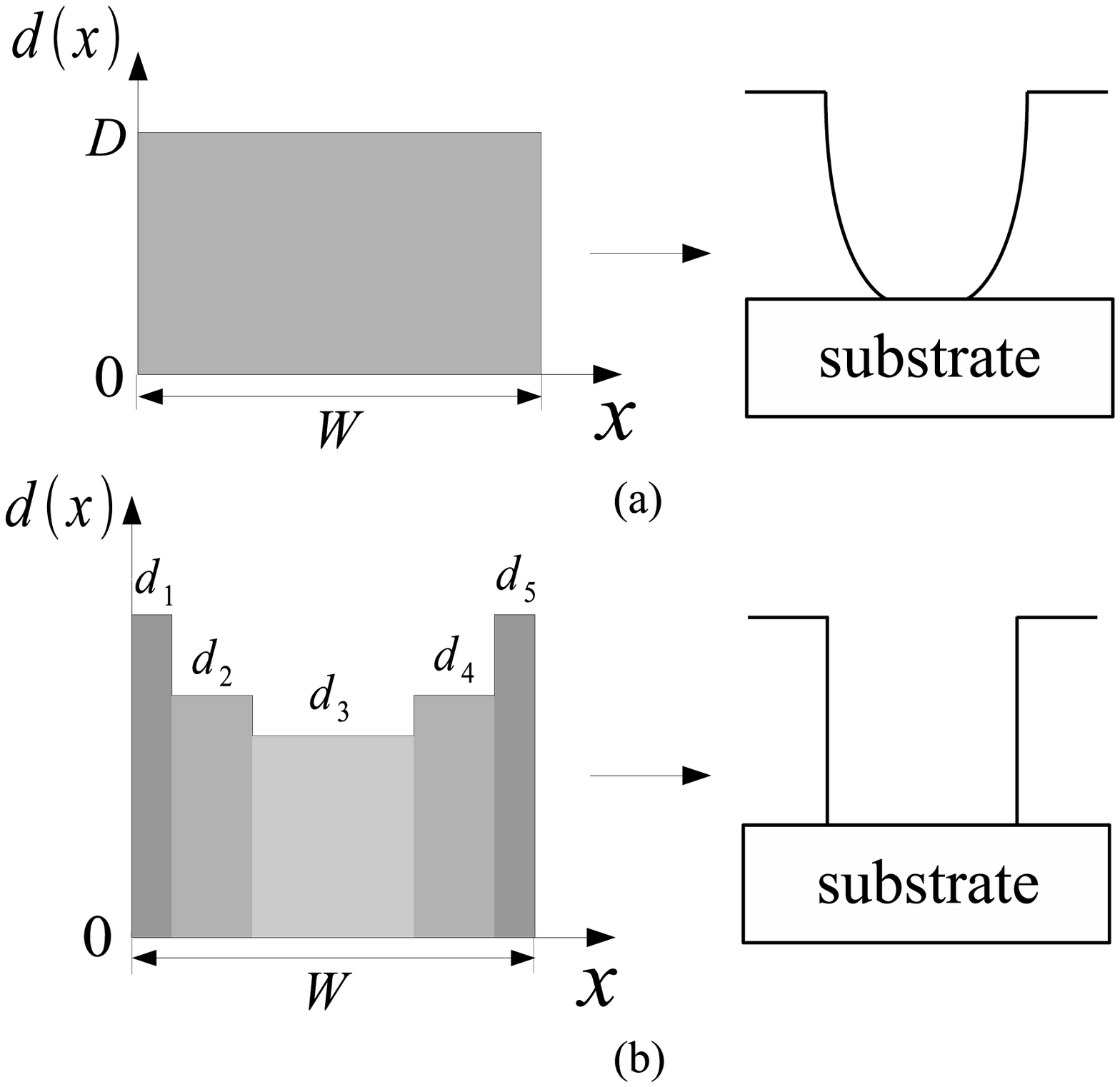} \\
\vspace{0.5cm}
 \caption{Dose distribution and the corresponding sidewall shape: (a) a uniform dose distribution and (b) the spatially-controlled dose distribution.}
\label{fig:totalDose}
\end{figure}

In order to avoid an impractically long computation time, the line is partitioned into $n$ regions along the length dimension as shown in Fig. \ref{fig:totalDose} and a dose for each region is to be determined.  That is, the solution from the optimization is a dose set $(d_1,d_2,\cdots,d_n)$ where $d_j$ is the dose for the $j$th region.  A fundamental difficulty of this optimization is that the optimal dose for a region has conflicts among layers, i.e., the dose required for a layer may be different from that for another layer. Also, the optimal dose for a region depends on the doses of the other regions.
In this study, the general-purpose optimization method of Simulated Annealing (SA) is adopted, which perturbs the doses of multiple regions in each iteration to find a globally optimal solution.

\begin{figure}[!htb]
 \centering
 \epsfxsize=12cm
 \ \epsfbox{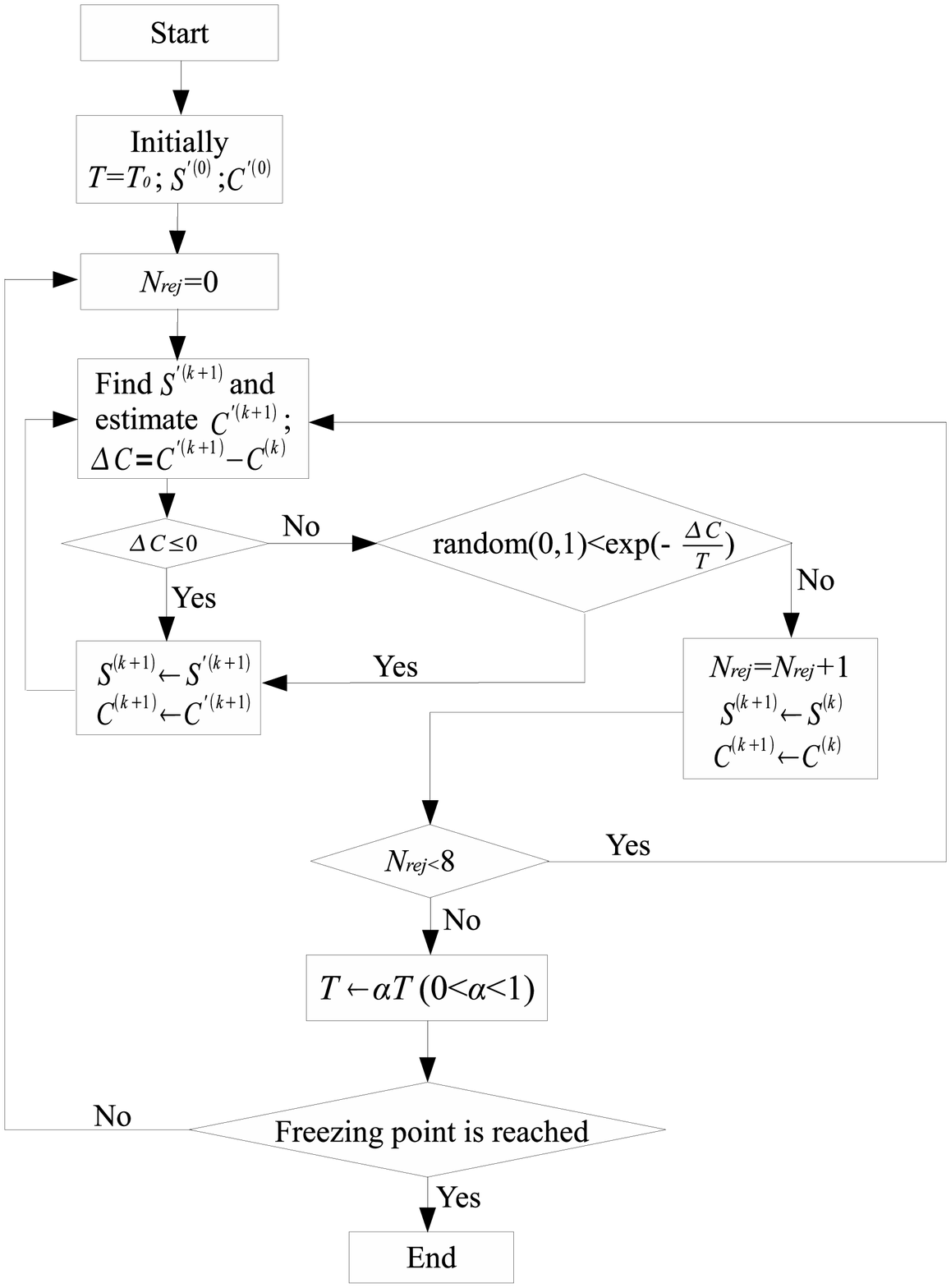} \\
 \caption{The flowchart of SA (simulated annealing) process}
\label{fig:SA}
\end{figure}

A single dose for all regions, minimizing the cost function, is first determined as an initial solution.  For evaluation of the cost function, the exposure distribution in the cross-section is computed through the convolution between the dose distribution and the point spread function.  Then, the resist development simulation is carried out to measure the dimensional errors in terms of line widths, i.e., $|rx_i-px_i|$.  The main optimization procedure of SA starts from the initial solution and iteratively derives the optimal or an acceptable solution.  The flowchart of SA is given in Fig. \ref{fig:SA} and the steps in SA are described below.  The solution obtained in the $k$th iteration is denoted by $S'^{(k)}$ = $(d^{(k)}_1, d^{(k)}_2,...,d^{(k)}_i,...,d^{(k)}_n)$ where $d^{(k)}_i$ is the dose for the $i$th region, derived in the $k$th iteration.

\begin{figure}[!htb]
 \centering
 \epsfxsize=12cm
 \ \epsfbox{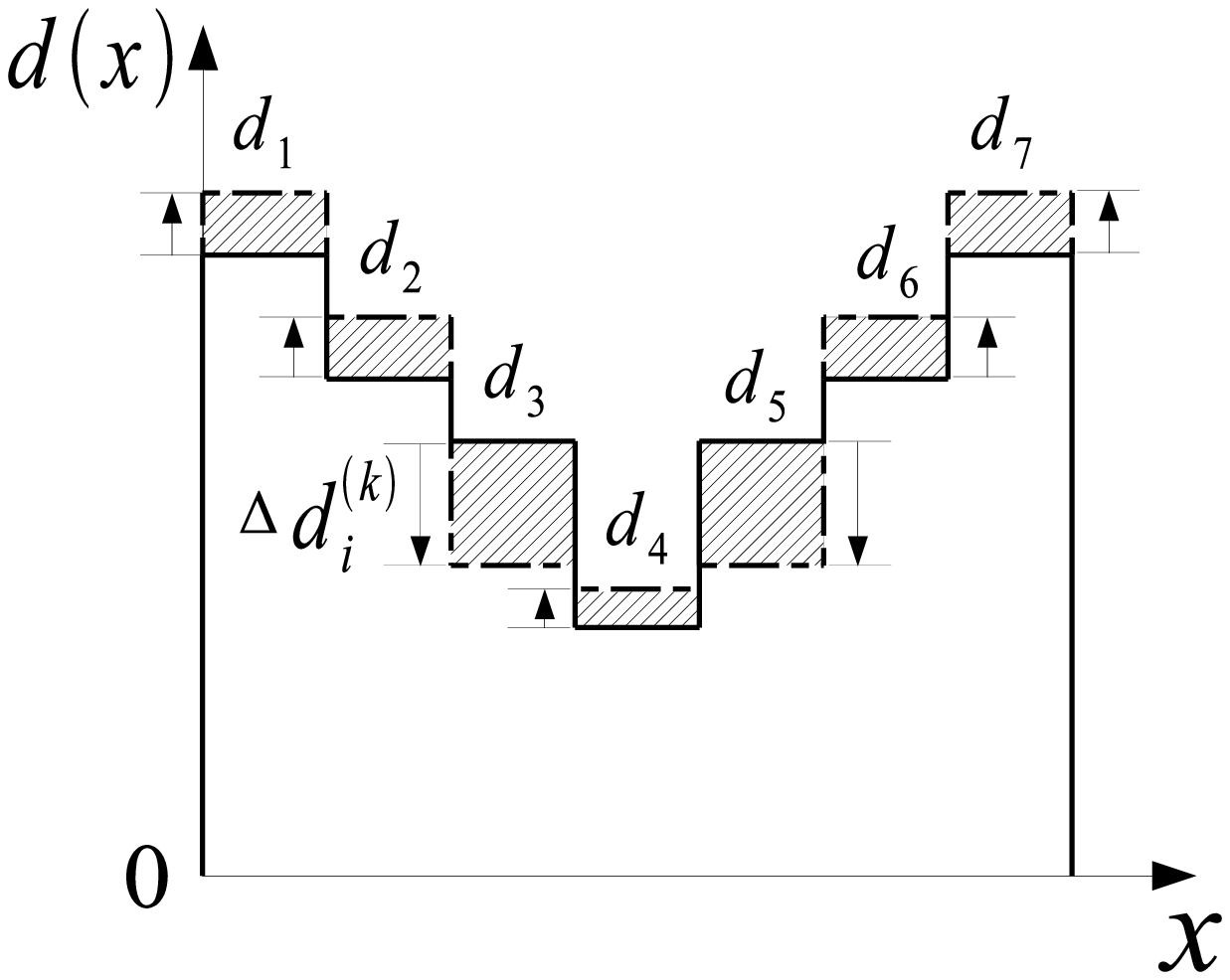} \\
 \caption{During the SA process, the doses of all regions of a line feature are adjusted.  The solid line and dashed lines represent the dose distribution before and after dose adjustment.
}
\label{fig:doseChange}
\end{figure}

\begin{figure}[!htb]
 \centering
 \epsfxsize=8cm
 \ \epsfbox{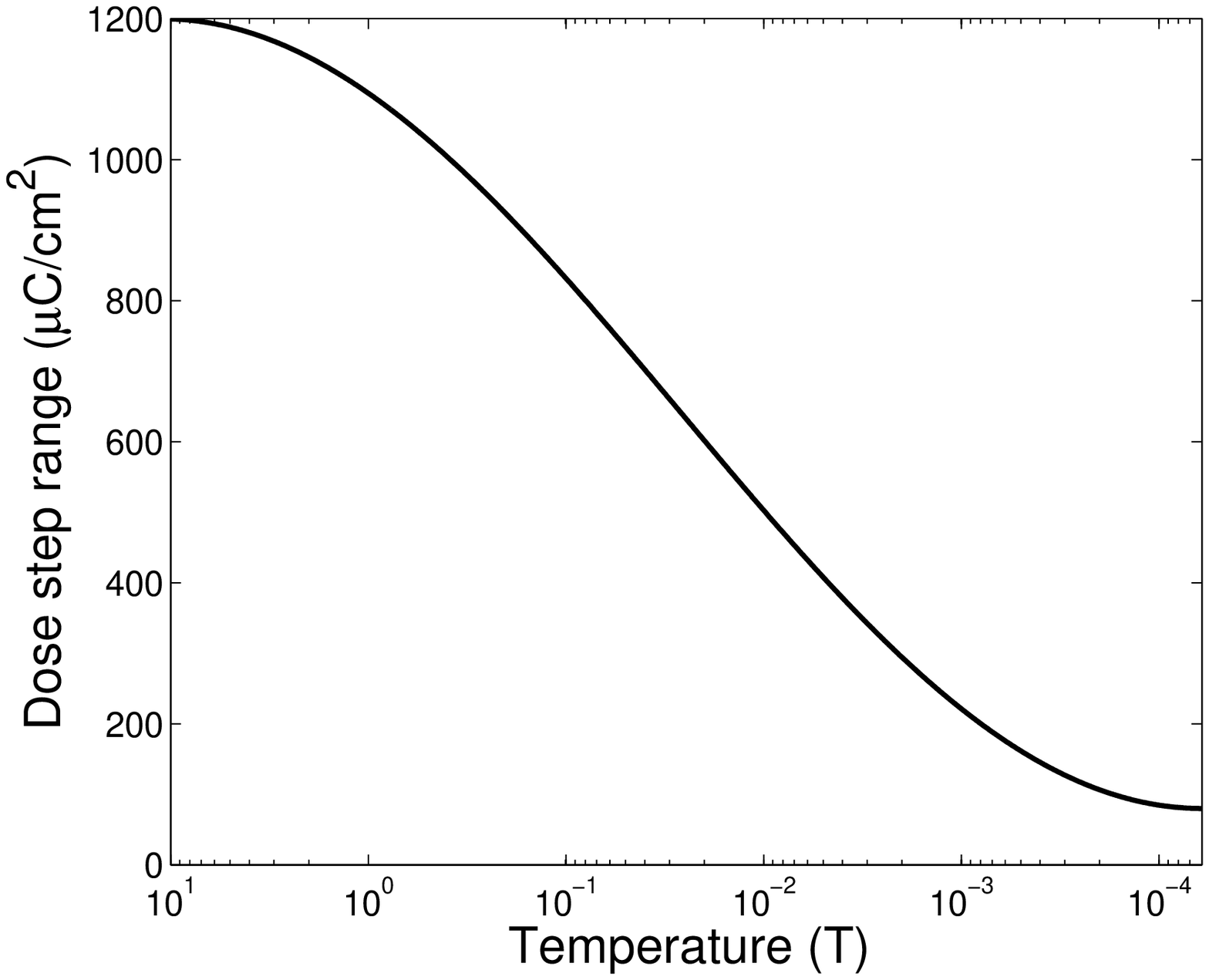} \\
 \caption{Dose step range vs temperature in the SA process.
}
\label{fig:doseStep}
\end{figure}

\begin{itemize}
\item[{\it Step 1}:]
Initially, the solution $S$ is set to $S'^{(0)}$=$($$d^{(0)}_1, d^{(0)}_2,...,d^{(0)}_i,...,d^{(0)}_n$$)$, and the temperature $T$ to a high value of $T_0$.  A possible initial dose distribution is a uniform distribution, i.e., $d^{(0)}_i=d^{(0)}_{j}$ for all $i,j$.  The cost function $C=max_i(|rx_i-px_i|)$ is evaluated based on $S'^{(0)}$ through resist-development simulation.  Let $C'^{(0)}$ denote the value of the cost function for $S'^{(0)}$.

\item[{\it Step 2}:]
Randomly perturb the current solution (spatial dose distribution) $S'^{(k)}$ to a potential new solution $S'^{(k+1)}$ = $S^{(k)}$+$(\Delta d^{(k+1)}_1,..., \Delta d^{(k+1)}_i,..., \Delta d^{(k+1)}_n)$, where $S^{(k)}$ is the accepted dose distribution in the $k$th iteration and $\Delta d^{(k+1)}_i$ is the amount of dose change for the $i$th region in the $(k+1)$th iteration.  Note that the doses of all regions are adjusted as illustrated in Fig. \ref{fig:doseChange}.  
In the case of single line, the dose distribution must be symmetric with respect to the center of line.  Therefore, only $\frac{n+1}{2}$ $\Delta d^{(k+1)}_i$'s need to be determined.  Determination of $\Delta d^{(k+1)}_i$ may be guided by a certain heuristic.  In this study, $\Delta d^{(k+1)}_i$ is computed as follows
\begin{equation}
\Delta d^{(k+1)}_i=\left(0.5\times(d_{max}-d_{min}-d_{minJump})\times(1+\cos(\frac{j\times\pi}{J}))+d_{minJump}\right)\times\left(r-0.5\right)
\label{eqn:T}
\end{equation}

where $d_{max}$ and $d_{min}$ are the upper and lower limits of dose allowed, $d_{minJump}$ is the minimum dose step of $\Delta d^{(k+1)}_i$, $j$ is the index for the $j$th temperature decrement from $T_0$ to the current $T$, $J$ is the total number of temperature decrements and $r$ is a random number ranging $[0,1]$.

Note that the dose step range is adjusted (decreased) as the temperature is decreased as shown in Fig. \ref{fig:doseStep}.  The cost function $C$ is evaluated for $S'^{(k+1)}$ to obtain its cost $C'^{(k+1)}$.

\item[{\it Step 3}:]
When $\Delta C=C'^{(k+1)}-C^{(k)}<0$ where $C^{(k)}$ is the value of cost function for $S^{(k)}$, $S'^{(k+1)}$ is accepted to become $S^{(k+1)}$.  If $\Delta C > 0$, $S'^{(k+1)}$ is still accepted with the probability of $exp(-\frac{\Delta C}{T})$.  This acceptance of a worse solution enables the hill-climbing capability of SA toward the globally optimal solution.  Otherwise, $S'^{(k+1)}$ is rejected in which case $S^{(k)}$ becomes $S^{(k+1)}$.  If the number of successive rejections $N_{rej}$ exceeds a certain threshold, go to Step 4.  Otherwise, go to Step 2.

\item[{\it Step 4}:]
The temperature $T$ is lowered according to $T \leftarrow \alpha\times T$ where $0<\alpha<1$.  That is, as the SA progresses, a worse solution is accepted less (since it is likely that the current solution is closer to the optimal solution).  Go to Step 2 if $T$ is above the final temperature.  Otherwise, go to Step 5.

\item[{\it Step 5}:]
The current solution is taken as the final solution (dose distribution).
\end{itemize}

\subsection*{Constraints}

The optimization of the dose distribution may be done with or without constraints such as total dose, developing time, etc.  It is always desirable to minimize the time to expose a pattern from the viewpoint of throughput.  The exposing time is mainly proportional to the total dose to be given to the pattern.  Also, the smaller the total dose is, the lower the charging effect is.  Hence, a dose distribution of which the total dose is smaller is better as long as it achieves an equivalent quality of the resist profile.  In most of our study, the constraint of the same total (average) dose was imposed, i.e., $D W=\int_0^W d(x) dx$ where $D$ is a certain dose level and $W$ is the line width as shown in Fig. \ref{fig:totalDose}.  In other words, the same total dose is redistributed over the line (feature) through optimization in order to achieve a certain target profile.

\section{Results and Discussion}
\label{sec:RD}

\begin{figure}[!htb]
 \centering
 \epsfxsize=8cm
 \ \epsfbox{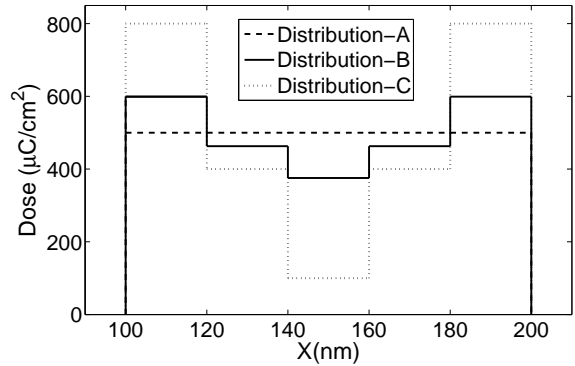} \\
 \caption{Three dose distributions of {\em Distribution-A}, {\em Distribution-B} and {\em Distribution-C}
}
\label{fig:doseDistribution}
\end{figure}

Three different dose distributions were considered in both simulation and experiment as shown in Fig. \ref{fig:doseDistribution}.  One, referred to as {\em Distribution-A}, is a uniform distribution.  Another, referred to as {\em Distribution-B}, is the one where the edge dose is moderately larger than the center dose.  The other, referred to as {\em Distribution-C}, has the edge dose much larger than the center dose.


\subsection{Simulation Results}

The test feature used in this study is a single line where the width and length of the line are 100 nm and 50 $\mu$m, respectively. The substrate system is composed of 300 nm poly (methyl methacrylate) (PMMA) on Si and the beam energy is assumed to be 50 KeV. The total (or average) dose is fixed in each set of results unless specified otherwise.

\begin{figure}[!htb]
 \centering
 \epsfxsize=8cm
 \ \epsfbox{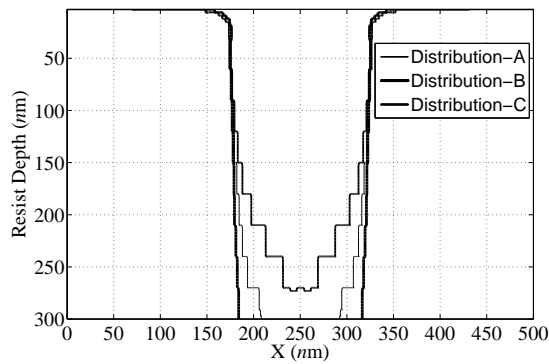} \\
 \caption{The remaining resist profiles (sidewall shapes) of {\em Distribution-A}, {\em Distribution-B} and {\em Distribution-C}.
}
\label{fig:profile}
\end{figure}

The relationship among the dose distribution, total dose, and developing time in terms of their effects on the sidewall shape has been analyzed through simulation.  The sidewall shapes obtained for the three different dose distributions (Fig. \ref{fig:doseDistribution}) with the total dose and developing time fixed are shown in Fig. \ref{fig:profile} and the line-width measurements are provided in Table \ref{tbl:doseDistri}.  It can be seen that a different dose distribution leads to a different sidewall shape.  For a vertical sidewall, the Distribution-B minimizes the width error (among the three).  The developing time required for a target sidewall shape ($rx_1=155$nm, $rx_5=145$nm, $rx_{10}=135$nm) was derived for each of the dose distributions in Fig. \ref{fig:doseDistribution}.  The Distribution-B requires the shortest developing time to achieve the sidewall shape closest to the target one (refer to Table \ref{tbl:time}).  The three dose distributions were scaled by a certain factor to achieve the target sidewall shape and the results are provided in Table \ref{tbl:totalDose}.  The Distribution-B requires the lowest total (average) dose while minimizing the width error.

\renewcommand{\baselinestretch}{1.5}
\begin{table}[!htb]
\begin{center}
\tabcolsep 2.5pt
\normalsize
\begin{tabular}{|c|c|c|c|c|c|} \hline
  & &  & \multicolumn {3} {c|} {Resist Profile $(nm)$ }   \\
Dose Distribution & Average Dose & Developing Time & \multicolumn {3} {c|} {Line Width }   \\
\cline {4-6}
  & ($\mu C/cm^2$) & $(sec)$ & $px_1$ & $px_5$ & $px_{10}$  \\ \hline
A:Dashed curve & 500.0 & 40.0  & 150.0 & 137.0 & 87.9  \\ \cline{1-6}
B:Solid curve & 500.0 & 40.0   & 151.2 & 143.0 & 133.0 \\ \cline{1-6}
C:Dotted curve & 500.0 & 40.0  & 155.8 & 125.0 & 0.0  \\ \hline
\end{tabular}
 \caption{Effects of the dose distribution on the sidewall shape with the total (average) dose and developing time fixed.
}
\label{tbl:doseDistri}
\end {center}
\end{table}

\renewcommand{\baselinestretch}{1.5}
\begin{table}[!htb]
\begin{center}
\tabcolsep 2.5pt
\normalsize
\begin{tabular}{|c|c|c|c|c|c|} \hline
  & &  & \multicolumn {3} {c|} {Resist Profile $(nm)$ }   \\
Dose Distribution & Average Dose & Developing Time & \multicolumn {3} {c|} {Line Width }   \\
\cline {4-6}
  & ($\mu C/cm^2$) & $(sec)$ & $px_1$ & $px_5$ & $px_{10}$  \\ \hline
A:Dashed curve & 500.0 & 49.8  & 155.8 & 148.0 & 132.0  \\ \cline{1-6}
B:Solid curve & 500.0 & 41.1   & 154.0 & 145.0 & 135.0 \\ \cline{1-6}
C:Dotted curve & 500.0 & 54.0  & 163.5 & 149.0 & 123.0  \\ \hline
\end{tabular}
 \caption{The developing time required to achieve the same (equivalent) sidewall shape with the total (average) dose fixed.
}
\label{tbl:time}
\end {center}
\end{table}

\renewcommand{\baselinestretch}{1.5}
\begin{table}[!htb]
\begin{center}
\tabcolsep 2.5pt
\normalsize
\begin{tabular}{|c|c|c|c|c|c|} \hline
 & &  & \multicolumn {3} {c|} {Resist Profile $(nm)$ }   \\
Dose Distribution & Average Dose & Developing Time & \multicolumn {3} {c|} {Line Width }   \\
\cline {4-6}
  & ($\mu C/cm^2$) & $(sec)$ & $px_1$ & $px_5$ & $px_{10}$  \\ \hline
A:Dashed curve & 630.0 & 40.0  & 156.4 & 149.0 & 134.0  \\ \cline{1-6}
B:Solid curve & 515.0 & 40.0   & 153.3 & 145.0 & 135.0 \\ \cline{1-6}
C:Dotted curve & 700.0 & 40.0  & 165.6 & 151.0 & 133.0  \\ \hline
\end{tabular}
 \caption{The total (average) dose required to achieve the same (equivalent) sidewall shape with the developing time fixed.
}
\label{tbl:totalDose}
\end {center}
\end{table}

\begin{figure}[!htb]
\centering
\epsfxsize=7.2cm
\ \epsfbox{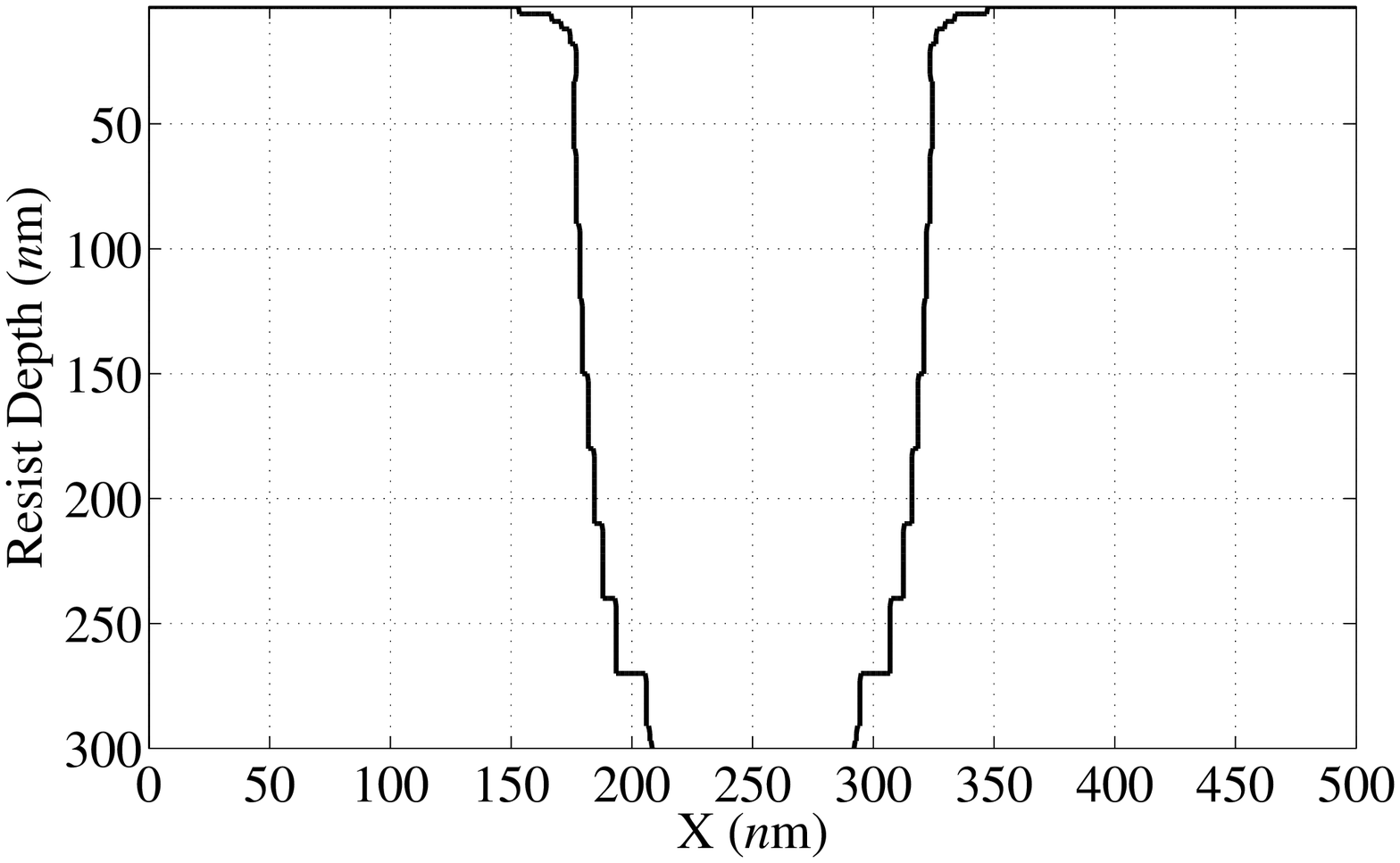} \hspace{1cm}
\epsfxsize=7.2cm
\ \epsfbox{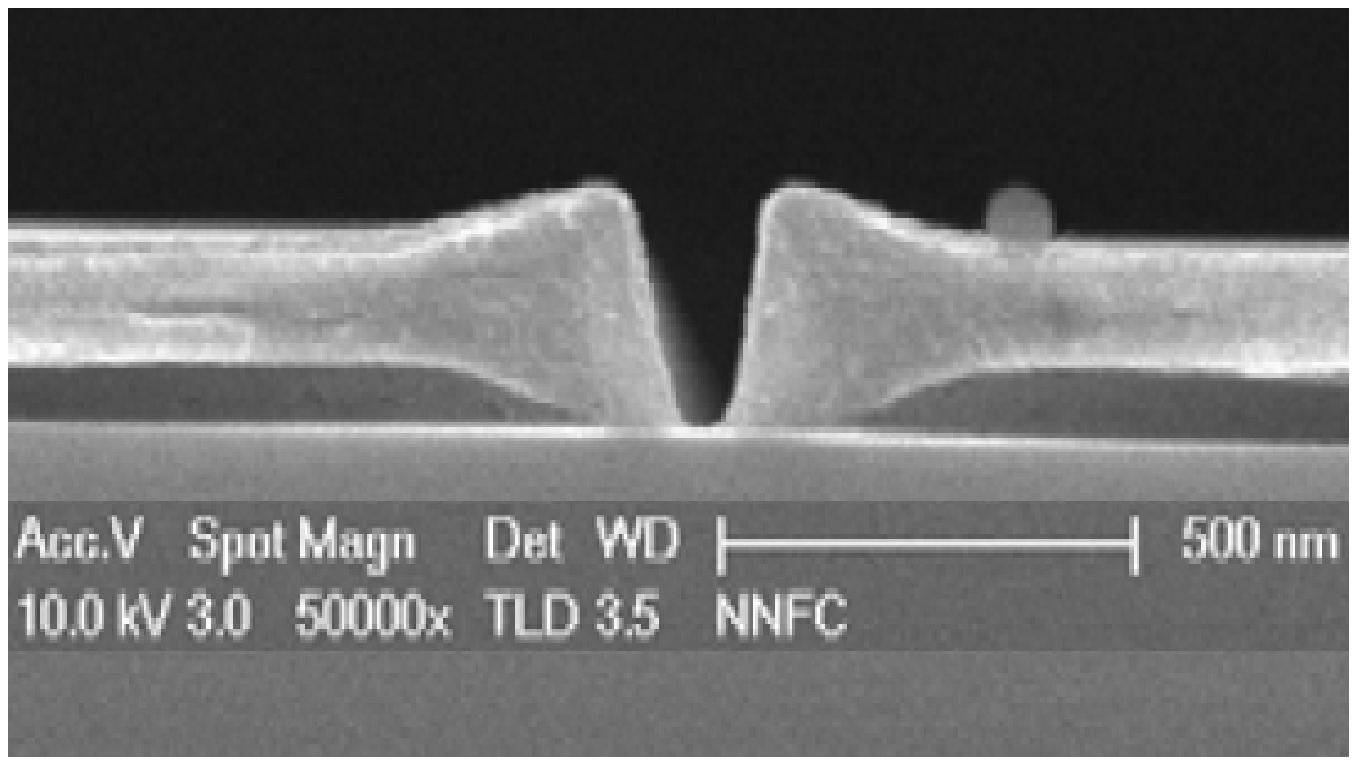} \\
\vspace{0.2cm}
(a) \hspace{7.5cm} (b) \\
\vspace{0.2cm}
\epsfxsize=7.2cm
\ \epsfbox{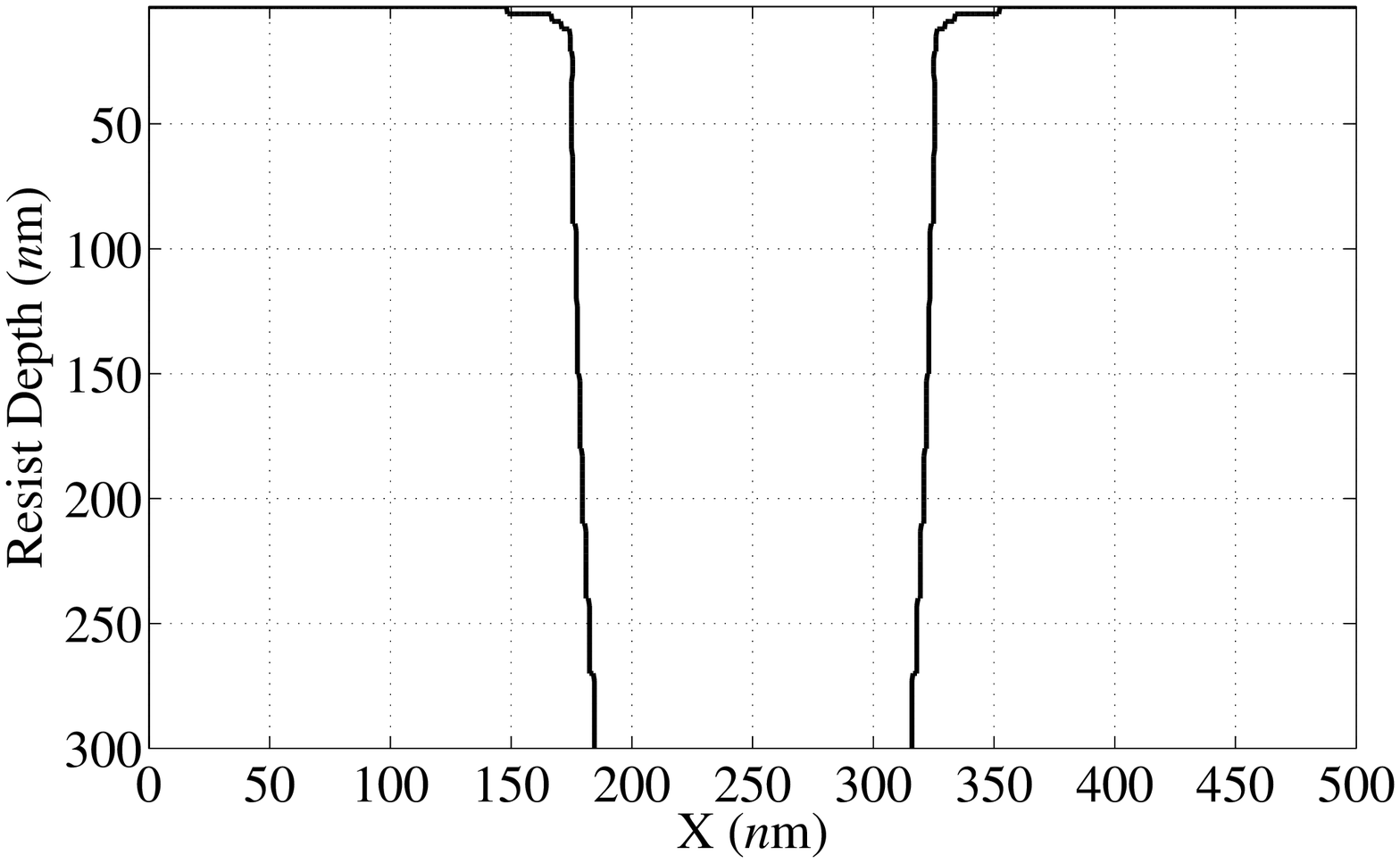} \hspace{1cm}
\epsfxsize=7.2cm
\ \epsfbox{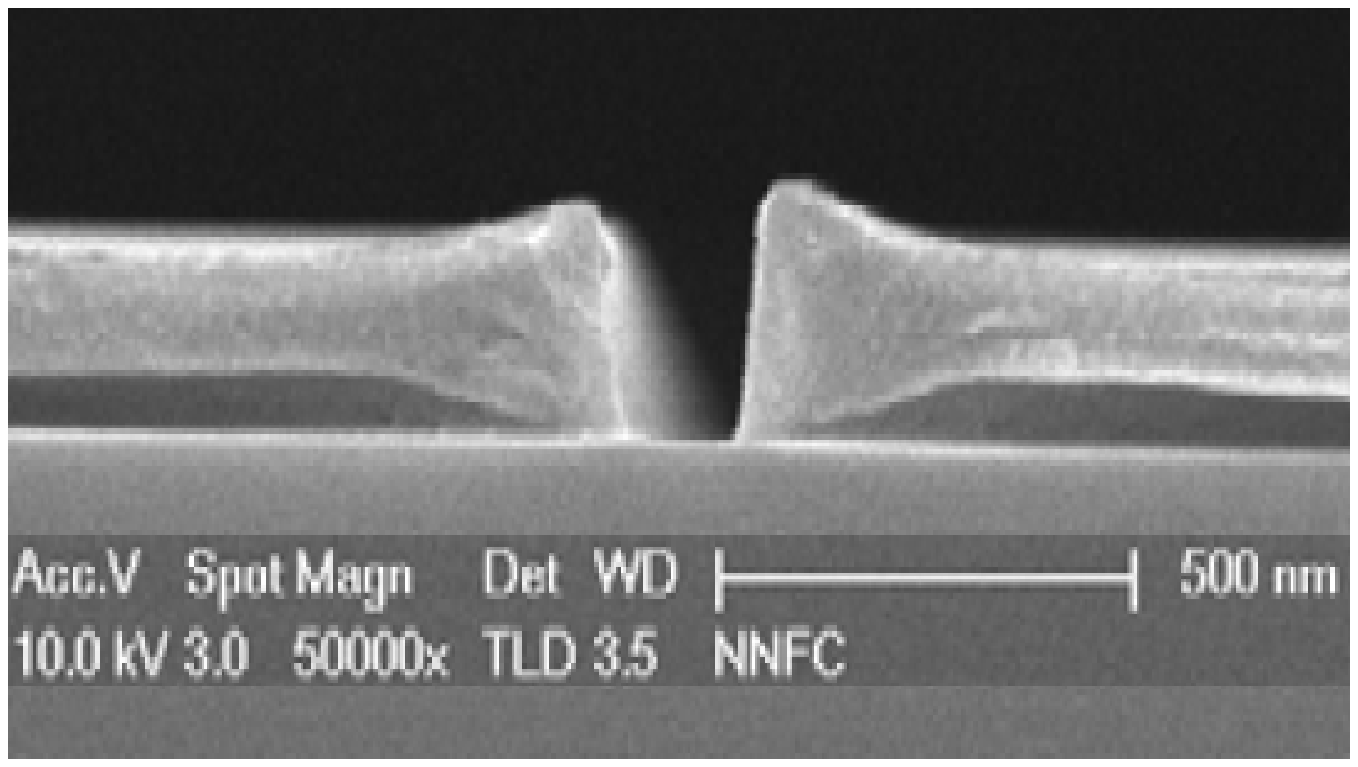} \\
\vspace{0.2cm}
(c) \hspace{7.5cm} (d) \\
\vspace{0.2cm}
\epsfxsize=7.2cm
\ \epsfbox{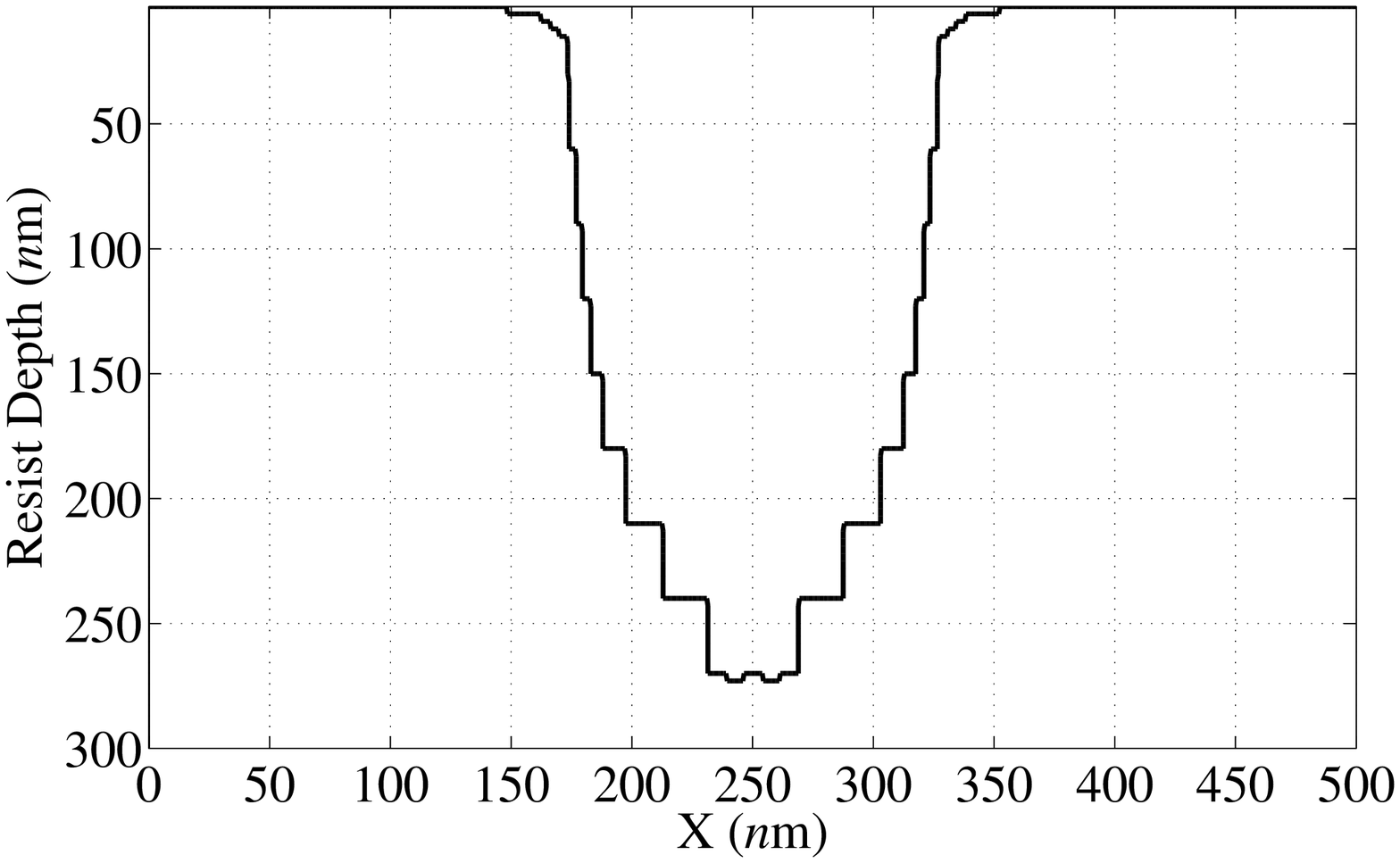} \hspace{1cm}
\epsfxsize=7.2cm
\ \epsfbox{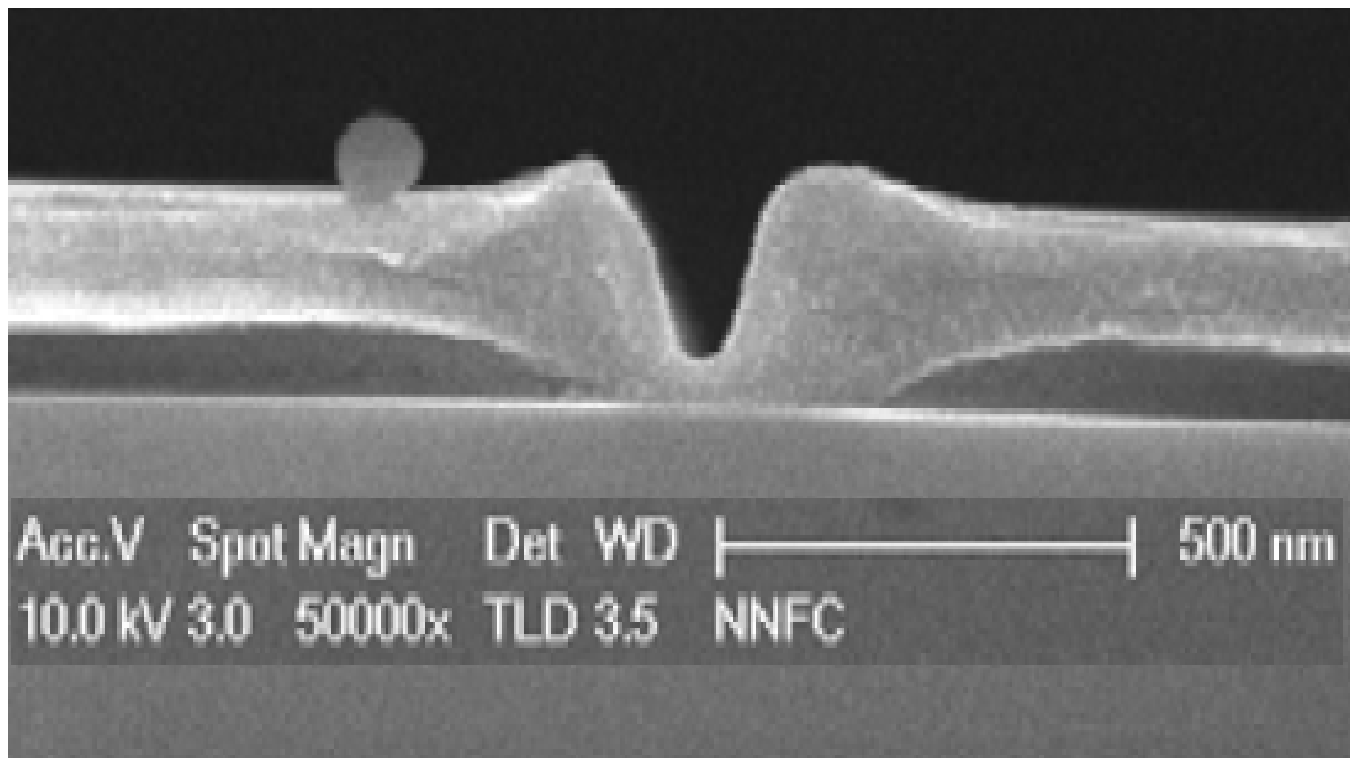} \\
\vspace{0.2cm}
(e) \hspace{7.5cm} (f) \\
\vspace{0.2cm}
\caption{Simulation and experimental results with the same average dose 500 $\mu C/cm^2$ (a) simulation result for {\em Distribution-A} (b) experimental result for {\em Distribution-A} (c) simulation result for {\em Distribution-B} (d) experimental result for {\em Distribution-B} (e) simulation result for {\em Distribution-C} (f) experimental result for {\em Distribution-C} where developing time: 40 $sec$, MIBK:IPA=1:2, 300 nm PMMA on Si (50 KeV).
}
\label{fig:data_500}
\end{figure}

In Fig. \ref{fig:data_500}, the remaining resist profiles obtained from the three different types of spatial dose distributions are provided for the average dose of 500 $\mu C/cm^2$.  The target sidewall shape is vertical.  When the dose is not controlled, i.e., a constant dose of 500 $\mu C/cm^2$ (Distribution-A), the sidewall shapes obtained are of overcut as can be seen in Fig. \ref{fig:data_500}(a), which is significantly different from the target one.  The sidewall shape obtained with spatial dose control (Distribution-B), shown in Fig. \ref{fig:data_500}(c), is much closer to the target sidewall shape. With a constant dose, developing rates in edge regions of the line are smaller than that at the center region, so the resist in edge regions is developed slower vertically, leading to an overcut.  When the dose is higher in edge regions of the line than in the center region as in the Distribution-B, the developing rate is higher in edge regions which causes lateral development at lower layers to start earlier. And also the exposure in unexposed regions tends to increase with depth. Therefore, lateral development following vertical development in the edge region catches up with vertical development right outside the edge region, leading to a more vertical sidewall shape. However, if the edge dose is increased beyond a certain level (with the average dose fixed) as in the Distribution-C, the effective developing rate is decreased significantly. This is due to the fact that the edge developing rate is much higher than that in its surrounding regime. Hence, the sidewall shape becomes overcut as seen in Fig. \ref{fig:data_500}(e).

\begin{figure}[!htb]
\centering
\epsfxsize=7.2cm
\ \epsfbox{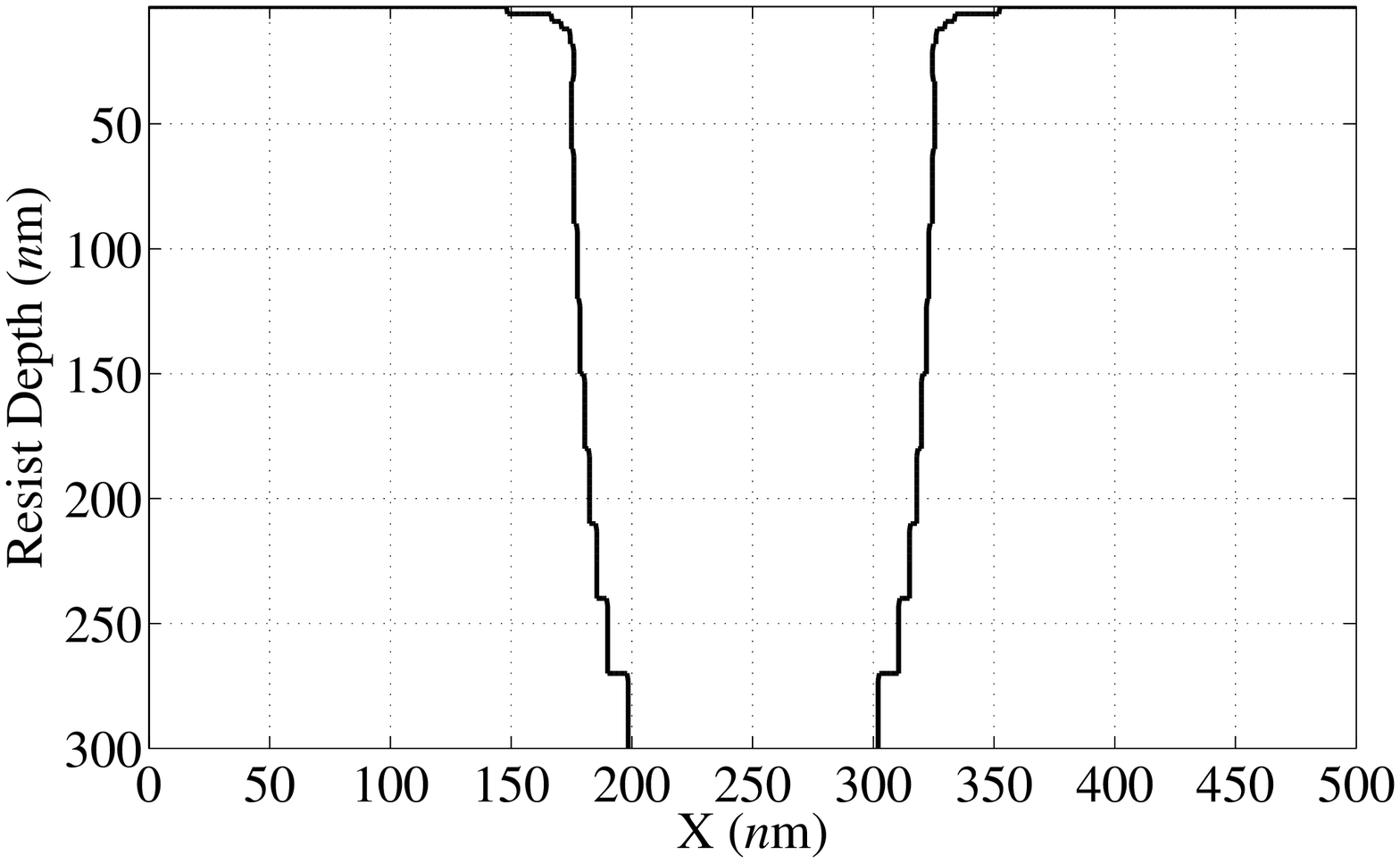} \hspace{1cm}
\epsfxsize=7.2cm
\ \epsfbox{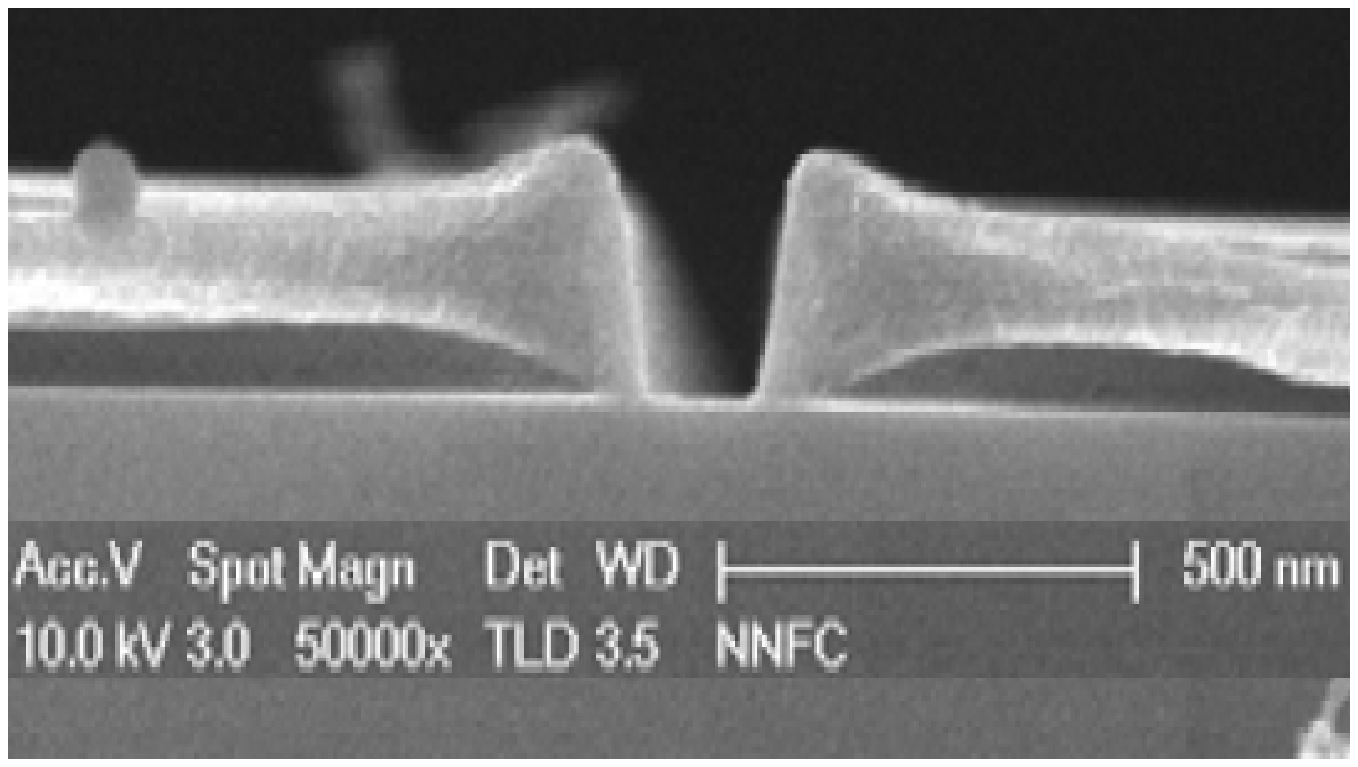} \\
\vspace{0.2cm}
(a) \hspace{7.5cm} (b) \\
\vspace{0.2cm}
\epsfxsize=7.2cm
\ \epsfbox{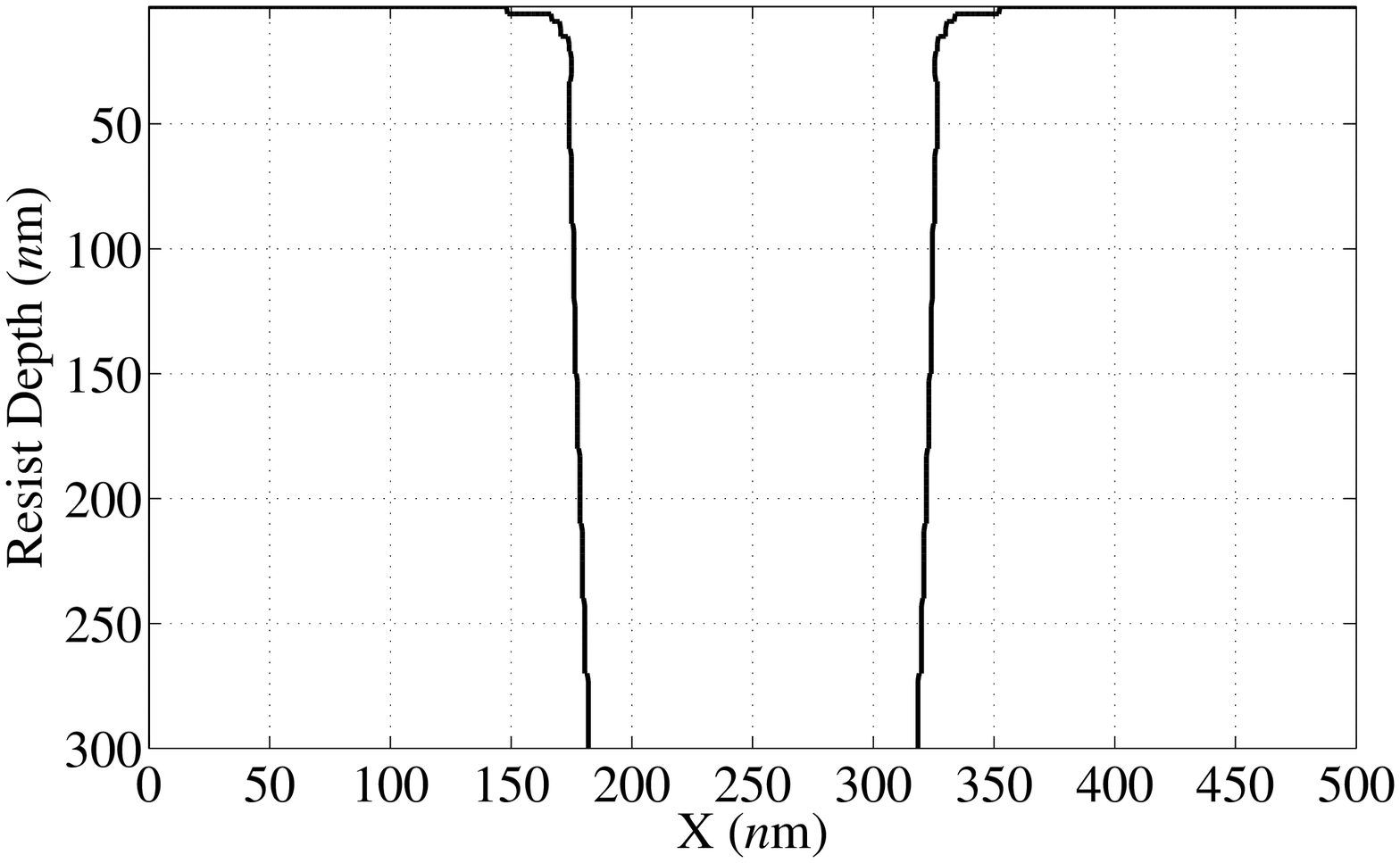} \hspace{1cm}
\epsfxsize=7.2cm
\ \epsfbox{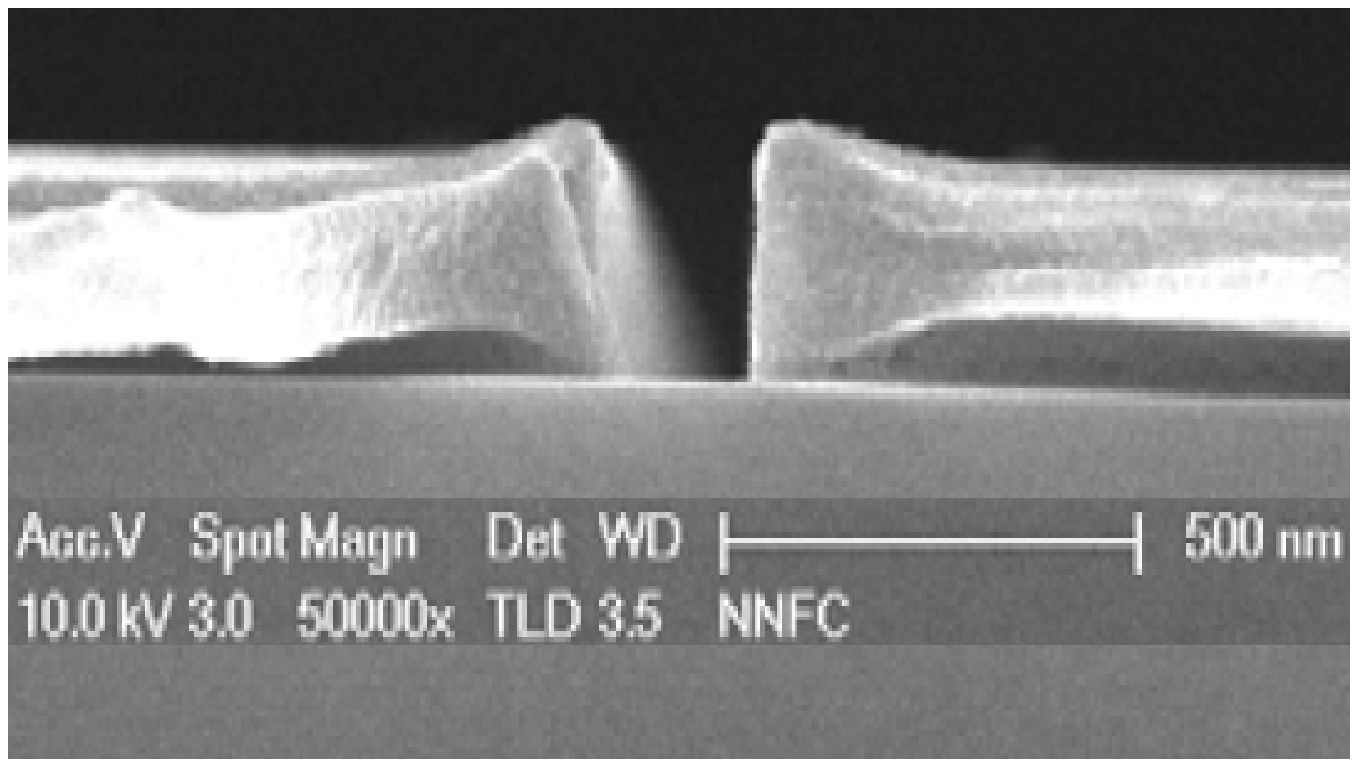} \\
\vspace{0.2cm}
(c) \hspace{7.5cm} (d) \\
\vspace{0.2cm}
\epsfxsize=7.2cm
\ \epsfbox{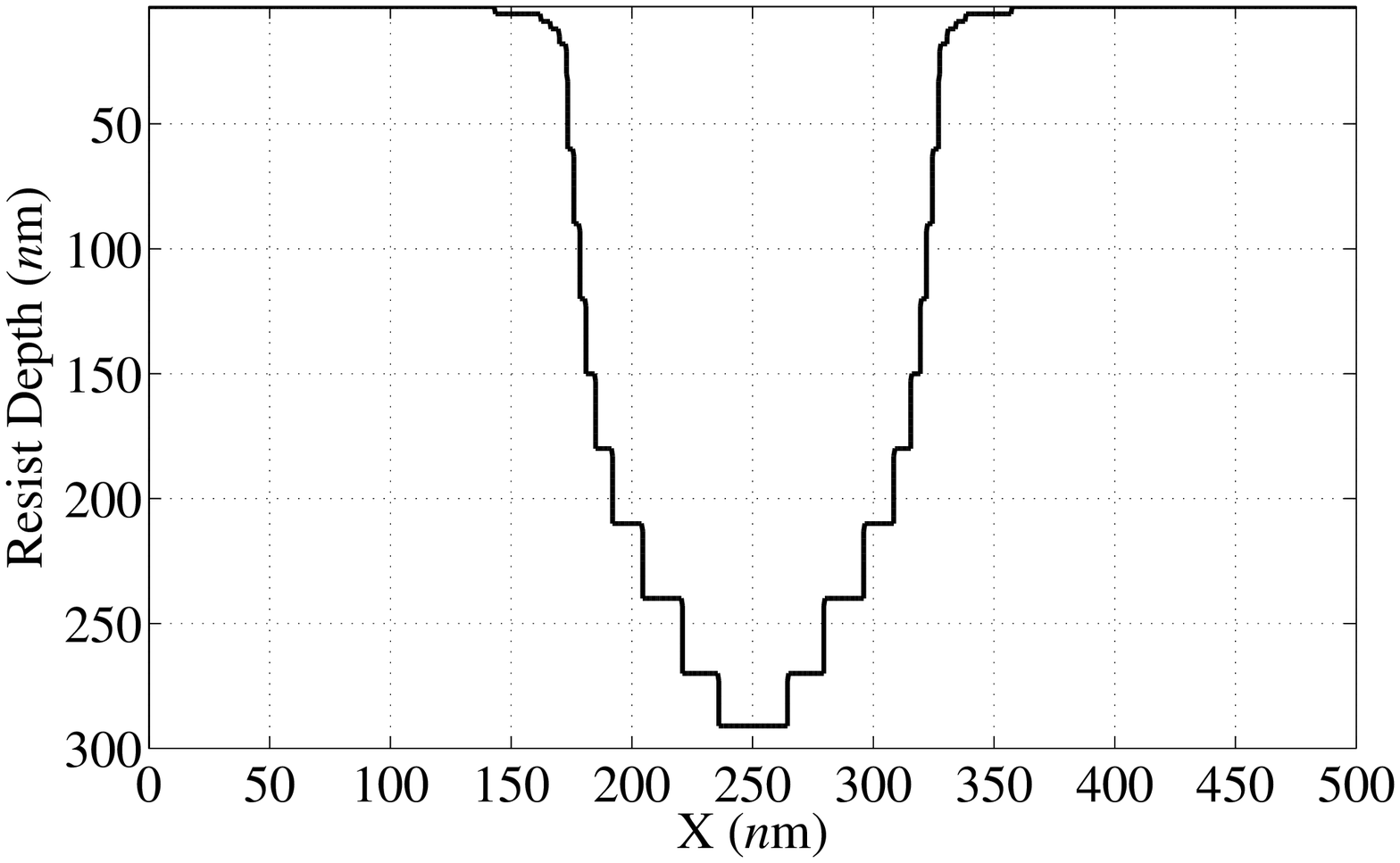} \hspace{1cm}
\epsfxsize=7.2cm
\ \epsfbox{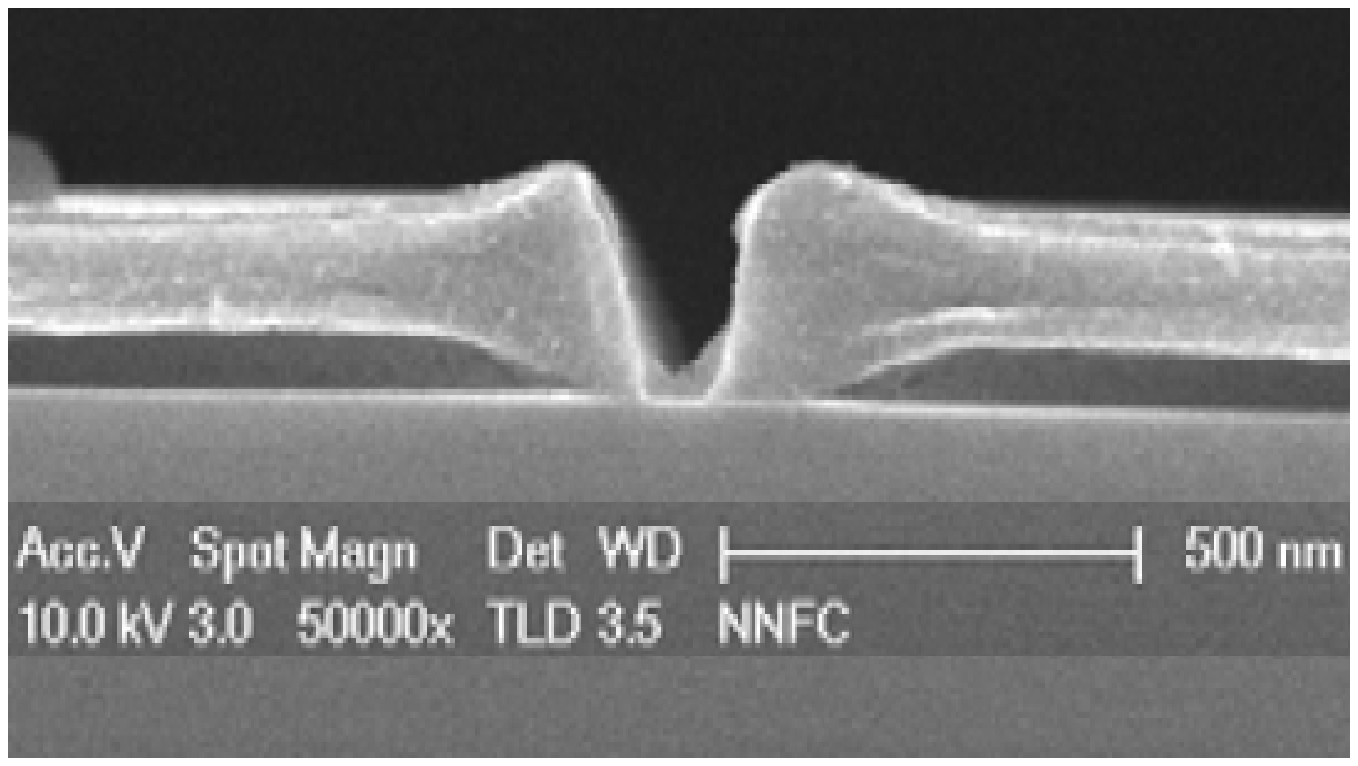} \\
\vspace{0.2cm}
(e) \hspace{7.5cm} (f) \\
\vspace{0.2cm}
\caption{Simulation and experimental results with the same average dose 525 $\mu C/cm^2$ (a) simulation result for {\em Distribution-A} (b) experimental result for {\em Distribution-A} (c) simulation result for {\em Distribution-B} (d) experimental result for {\em Distribution-B} (e) simulation result for {\em Distribution-C} (f) experimental result for {\em Distribution-C} where developing time: 40 $sec$, MIBK:IPA=1:2, 300 nm PMMA on Si (50 KeV).
}
\label{fig:data_525}
\end{figure}

The same set of the results for the average dose of 525 $\mu C/cm^2$ are provided in Fig. \ref{fig:data_525}.  The sidewall for the increased constant dose, in Fig. \ref{fig:data_525}(a), is more vertical than that in  Fig. \ref{fig:data_500}(a), but still not so vertical as that in Fig. \ref{fig:data_500}(c).  As in the case of the average dose of 500 $\mu C/cm^2$, the spatial dose control with a small dose difference between the edge and center regions results in a much more vertical sidewall as can be seen in Fig. \ref{fig:data_525}(c), which is almost the same as that in Fig. \ref{fig:data_500}(c).  Again, too large a dose difference between the edge and center regions leads to an overcut shown in Fig. \ref{fig:data_525}(e).

\subsection{Experimental Results}

The single line was fabricated with the three different types of dose distributions considered in the simulation. The substrate system was prepared by spin-coating a Si wafer with 300 nm PMMA and soft-baked at 160$^o~C$ for 1 minute.  The structure was written using an Elionix ELS-7000 e-beam tool with acceleration voltage of 50 KeV and beam current of 100 pA.  The sample was developed in MIBK:IPA$~$=$~$1:2 for 40 seconds. The remaining resist was coated with 10 nm Pt before the cross section was imaged by a FEI FE-SEM (Sirion).  For easier inspection of the cross section, the length of the line was increased to 500 $\mu$m.  The SEM images of the cross-section are provided in Fig. \ref{fig:data_500} and Fig. \ref{fig:data_525} with the average dose 500 $\mu C/cm^2$ and 525 $\mu C/cm^2$, respectively.

It has been experimentally verified that by controlling the spatial dose distribution one can achieve different shapes of sidewalls though the total amount of dose given to the line remains the same.  Specifically, the simulation result that the Distribution-B achieves the sidewall shape closest to the target one (vertical sidewall) is consistent with the experimental result (Fig. \ref{fig:data_500}(d) and Fig. \ref{fig:data_525}(d)).  Also, as the simulation result indicates, the edge dose should not be increased too much when the target sidewall is vertical as shown in Fig. \ref{fig:data_500}(f) and Fig. \ref{fig:data_525}(f).  In addition, it is possible to reduce the total dose while achieving an equivalent sidewall shape.

\section{Summary}
\label{sec:SM}

In this study, the problem of controlling the sidewall shape of the resist profile has been addressed. A practical scheme to achieve a given target sidewall of a line by controlling the e-beam dose spatially based on a 3-D exposure model is described.  A line is partitioned into regions along the length dimension and a dose is determined for each region.  In the proposed scheme, it is attempted to find the optimum dose distribution by the general-purpose optimization scheme, Simulated Annealing.  Through computer simulation, the effects of the factors such as dose distribution, total dose, and developing time, and performance of the dose control scheme have been analyzed.  The simulation results have been verified through experiments.  The current and future research efforts include application of the proposed scheme to more general patterns and further experimental verification.

\end{document}